\documentclass[sigconf]{acmart}
\AtBeginDocument{%
  }

\setcopyright{acmlicensed}
\copyrightyear{2018}
\acmYear{2018}
\acmDOI{XXXXXXX.XXXXXXX}
\acmConference[Conference acronym 'XX]{Make sure to enter the correct
  conference title from your rights confirmation email}{June 03--05,
  2018}{Woodstock, NY}
\acmISBN{978-1-4503-XXXX-X/2018/06}

\begin{document}

\author{Zaibei Li}
\affiliation{%
  \institution{University of Copenhagen}
  \city{Copenhagen}
  \country{Denmark}
}
\email{zali@di.ku.dk}

\author{Qiuchi Li}
\affiliation{%
  \institution{Beijing Institute of Technology}
  \city{Beijing}
  \country{China}
}
\email{liqiuchi@bit.edu.cn}

\author{Shunpei Yamaguchi}
\affiliation{%
  \institution{Hiroshima City University}
  \city{Hiroshima}
  \country{Japan}
}
\email{yamaguchi-s@hiroshima-cu.ac.jp}

\author{Daniel Spikol}
\affiliation{%
  \institution{University of Copenhagen}
  \city{Copenhagen}
  \country{Denmark}
}
\email{ds@di.ku.dk}

\settopmatter{printacmref=false} 
\setcopyright{none} 
\renewcommand\footnotetextcopyrightpermission[1]{} 
\pagestyle{plain}
\title{BadgeX: IoT-Enhanced Wearable Analytics
Meets LLMs for Collaborative Learning}


\renewcommand{\shortauthors}{Li et al.}

\begin{abstract}
We present BadgeX, a novel system integrating lightweight wearable IoT devices (smart badges/smartphones) with Large Language Models (LLMs) to enable real-time collaborative learning analytics. The system captures multimodal sensor data (e.g., audio, image, motion, depth) from learners, processes it into structured features, and employs an LLM-driven framework to interpret these features, generating high-level insights grounded in learning theory. A pilot study demonstrated the system's capability to capture rich collaboration traces and for an LLM to produce plausible, theoretically coherent narrative analyses from sensor-derived features. BadgeX aims to lower deployment barriers, making complex collaborative dynamics visible and offering a pathway for real-time support in educational settings.
\end{abstract}

\ccsdesc[500]{Human-centered computing}
\ccsdesc[300]{Ubiquitous and mobile computing}
\ccsdesc[100]{Empirical studies in ubiquitous and mobile computing}

\keywords{IoT, Wearable Devices, Collaboration Analytics, Large-Language Model}

\begin{teaserfigure}
  \includegraphics[width=\textwidth]{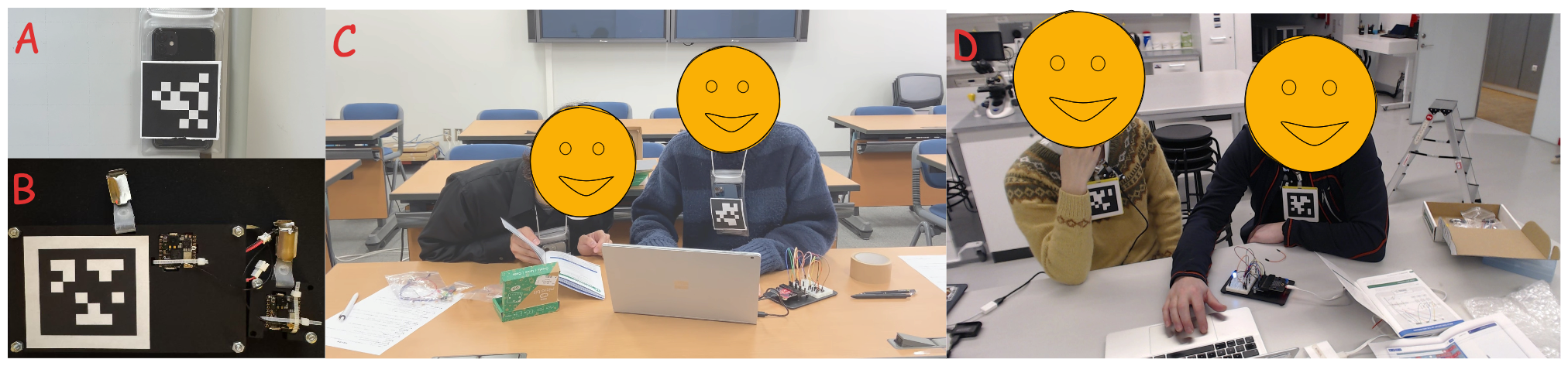}
  \caption{A, B are the wearable sensing devices; C, D are BadgeX in action}
  \label{fig:teaser}
\end{teaserfigure}

\received{20 February 2007}
\received[revised]{12 March 2009}
\received[accepted]{5 June 2009}

\maketitle

\section{Introduction}
Wearable and mobile sensing technologies are opening new frontiers in learning analytics by capturing rich, in situ data about student collaboration. In group learning activities, important processes like knowledge sharing, negotiation \& coordination, and team maintenance often go unobserved or unanalyzed in real time. Conventional multimodal learning analytics (MMLA \cite{Blikstein2016, ochoa2022multimodal, spikol_supervised_2018}) setups that use fixed sensors can be bulky and hard to deploy, limiting their adoption in everyday classrooms. BadgeX integrates lightweight wearable IoT devices with Large Language Models (LLMs) to address these gaps to support real-time collaborative learning analytics. The system collects multimodal signals from each learner through wearables (e.g., smart badges or smartphones) and translates low-level sensor data into high-level insights grounded in learning theory.

\section{Related Work}
\textit{Wearables for Collaboration Analytics}: Prior studies have explored using wearable sensors to analyze teamwork and collaboration. For example, sociometric badges, originating from MIT Media Lab \cite{wu_mining_2008}, use sound and RF signals for scalable and versatile interaction analysis, have evolved into OpenBadge \cite{lederman_openbadge} and Rhythm \cite{lederman_rhythm}. Hitachi’s Business Microscope \cite{wakisaka} and the Sensor-based Regulation Profiler \cite{yamaguchi_iot_2022}, both using business-card-sized sensors for workplace behavior analysis. These wearable devices continuously and unobtrusively log group interaction patterns, reducing the need for manual observation. Such research affirms that wearable technology enables continuous onsite data collection from multiple people, yielding new insights into group dynamics and engagement. However, many existing MMLA solutions require complex instrumentation or specialized hardware, making them less feasible for everyday classroom use. BadgeX builds on this prior work by using lightweight ubiquitous devices to lower deployment barriers while upholding data fidelity and diversity.

\textit{LLMs in Real-Time Analytics}: Parallel to sensor advances, large language models have emerged as powerful pattern interpreters. Recent research suggests that LLMs can enhance the analysis and interpretation of raw sensor data, by bringing contextual knowledge and flexible reasoning to bear on multimodal inputs. In education, early systems have begun to exploit LLMs for learning analytics – for instance, VizGroup \cite{tang_vizgroup_2024} visualizes students’ collaborative behavior and sends proactive alerts to instructors by leveraging LLM-based analysis of programming collaboration logs. Talk2Care \cite{yang_talk2care_2024} utilizes the LLMs to facilitate patient/provider inter-personal communication. WhiteHead \cite{whitehead_utilizing_2025} and Clayton \cite{cohn_chain--thought_2024} leverage the LLMs for automatic coding in learning analytics. These approaches hint at the potential of foundation models to serve as “brains” that synthesize low-level events into meaningful narratives or assessments for instructors and learners. BadgeX is novel in applying LLMs to wearable sensor streams in education, effectively merging IoT and AI: the IoT devices provide real-time data from the physical collaboration, and the LLM provides a high-level interpretation grounded in pedagogical constructs.

\section{BadgeX: Design and Prototype}
\subsection{Wearables and Sensing}
At the heart of BadgeX is a wearable sensing network composed of smartphones and/or custom smart badges worn by each student, as shown in Figure \ref{fig:teaser}. Initially, we customized our wearables with Arduino Nicla vision boards, then we chose smartphones as the sensing platform due to their high fidelity and rich sensor suites. In our prototype, each learner wears a smartphone on a lanyard (positioned on the chest with an AprilTag \cite{apriltag} - a fiducial marker for unique ID tracking), running a data collection app. This mobile IoT setup is lightweight, wireless, and deployable in real classrooms without elaborate infrastructure. The phone badge utilizes the microphone, camera, IMU, and LiDAR sensors to collect audio, image, motion, and depth data. Environmental fixed-position webcams (Figure \ref{fig:teaser}: C\&D) were deployed to capture images of participants’ physical interactions with each other and with the surrounding environment.

All sensor data is time-stamped and either stored locally or streamed over an ad hoc network to a base station (e.g., a laptop or single board computer) for real-time processing. The IoT data pipeline is designed to be efficient and real-time: sensors collect multimodal data and stream it directly to a distributed dedicated base station responsible for processing a specific data modality or to a centralized RTMP server, which the base station subscribes to. Raw sensor data is processed on base stations using AI services deployed on powerful local servers, with feature results synchronized according to their timestamps, which follows the edge-computing paradigm.

By decoupling data acquisition from processing, the system supports real-time analysis while remaining mobile, capable of running on battery-powered devices with limited bandwidth.

\subsection{LLM Framework: From Signals to Learning Construct}

\begin{figure*}[t]
  \centering
  \includegraphics[width=0.85\textwidth]{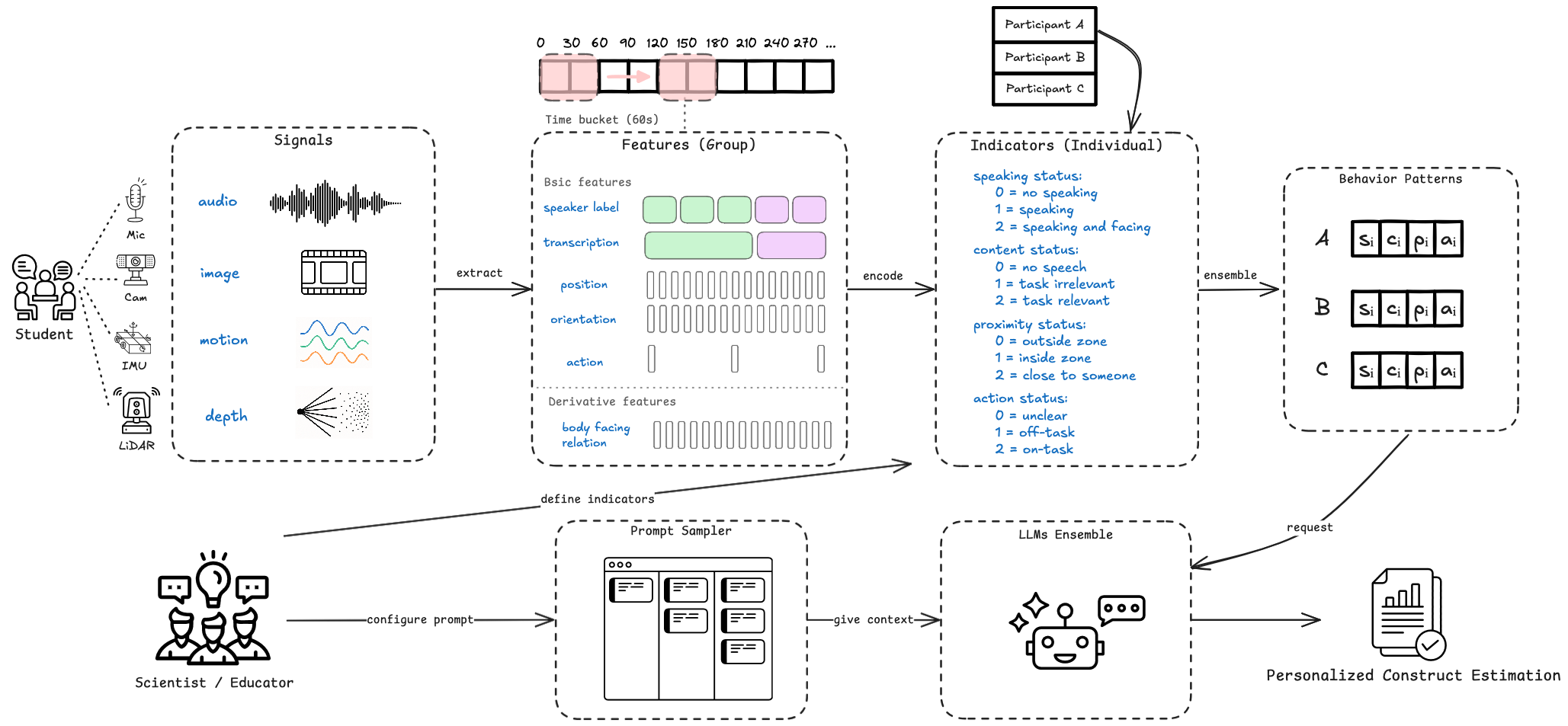}
  \caption{LLM-Enhanced IoT Analytical Framework of BadgeX: ubiquitous sensors capture multimodal signals, IoT pipelines extract and aggregate features, and LLMs interpret these features to generate high-level analytics on learning construct.}
  \label{analytical_framework}
\end{figure*}

BadgeX applies LLMs to translate the structured sensor data into high-level insights. In prior work, Xavier \cite{ochoa2022multimodal} proposed a dual-process mapping from multimodal data to learning constructs, while Yan \cite{yan2025} employed latent class analysis to integrate monomodal behavioural indicators into data-driven multimodal latent classes. Rather than hand-coding one-to-one rules, our framework (Figure \ref{analytical_framework}) feeds generic theory-based indicators plus construct- and experiment-specific context into an LLM, allowing the model’s generative reasoning to flexibly map diverse behavioural patterns to one or multiple learning constructs.

As a learning session begins, the system continuously extracts human-observable features from audio, image, motion, and depth data, each using modality-specific time windows, leveraging AI services (speaker recognition, speech to text, computer vision, VLMs scene understanding, etc). Those modality-specific features are aggregated into synchronized 60-second time buckets using a 30-second sliding window to capture evolving collaborative dynamics. For instance, speaker labels update continuously every 3 seconds. Speech transcriptions are generated dynamically, based on each speaker's uninterrupted speaking turn length. Positional and orientation data are recorded as discrete samples every 1 second, capturing participants' most recent states. Action recognition is performed at regular 30-second intervals, analyzing a small number of video frames captured at each timestamp from one or multiple camera perspectives. In addition to directly extracted sensor features, derived features, such as body-facing relationships, are computed by evaluating inter-person distances and orientation angles. Subsequently, these features are encoded into meaningful, categorical indicators grounded in established learning theory \cite{Giannakos23}. Encoding is performed through a combination of rule-based classification and semantic analyses provided by LLM/NLP techniques. Within each time bucket, each participant \( i \) has a behavioral pattern \( P_i = [s_i, c_i, p_i, a_i] \), representing speaking status (\( s_i \)), speech content (\( c_i \)), proximity (\( p_i \)), and action status (\( a_i \)), \(
\small
s_i=[s_{i,1},\dots,s_{i,d_s}]^\top\!\in\!\mathbb{R}^{d_s}\allowbreak,\;
c_i=[c_{i,1},\dots,c_{i,d_c}]^\top\!\in\!\mathbb{R}^{d_c}\allowbreak,\;
p_i=[p_{i,1},\dots,p_{i,d_p}]^\top\!\in\!\mathbb{R}^{d_p}\allowbreak,\;
a_i=[a_{i,1},\dots,a_{i,d_a}]^\top\!\in\!\mathbb{R}^{d_a}\allowbreak,\;
P_i=\operatorname{concat}(s_i,c_i,p_i,a_i)\!\in\!
      \mathbb{R}^{d_s+d_c+d_p+d_a}.
\) The group behavior patterns, represented as multimodal composite vectors, are combined with contextual information about target learning constructs defined by a prompt sampler. We use a few-shot prompting approach, providing the LLM with illustrative mappings from patterns to specific construct assessments. Essentially, the LLM is tasked with classifying or describing the state of each participant's learning constructs based on their sensor-derived evidence.

Leveraging the generative and reasoning capabilities of LLMs ensures flexibility, allowing the incorporation of general world knowledge and subtle contextual reasoning. Moreover, the framework is highly generalizable: by adjusting the prompts, the same analytical pipeline can address different constructs or collaboration skills without retraining specialized models.

During development, we explored different output formats from the LLM, including scoring constructs numerically and generating open-ended textual analyses. Ultimately, narrative explanations were preferred for the pilot because they provided richer, more actionable feedback. The resulting analyses can be displayed on a web dashboard for instructors to review post-session, or delivered immediately to students via a chatbot interface as real-time feedback.

\section{Pilot Study}
We piloted BadgeX in a controlled collaborative learning session to evaluate the full end-to-end pipeline. The session involved two participants engaged in a 43-minute STEM problem-solving task while wearing custom-designed Arduino-based smart badges. To guide our analysis, we adopted the collaborative problem-solving (CPS) framework from \cite{sun_towards_2020}, which defines key facets of CPS including: constructing shared knowledge, negotiation and coordination, and maintaining team function.

Setup: the audio feature extraction pipeline includes a denoising algorithm \cite{denoiser}, Silero voice activity detection \cite{silero}, and Titanet speaker embeddings \cite{titanet} for speaker recognition, and WhisperX \cite{bain2022whisperx} for speech transcription. For visual analysis, we employed gaze detection \cite{ryan2025gazelle} and AprilTag-based identity alignment, as well as LLM (gemini-2.5-pro-exp \cite{gemini}) for action recognition. For spatial tracking, we used a visual-inertial odometry algorithm \cite{zhang18} to estimate position and orientation, with AprilTag detection as an alternative.

Data Capture Results: The system successfully captured a rich set of multimodal collaboration traces. Audio was manually transcribed, and video frames were labeled for action recognition to provide ground truth. Our automated speech pipeline yielded a diarization error rate of 17.8\% and a word error rate of 26.4\%. For action recognition, 161 out of 176 predictions matched human annotations, demonstrating 90\% alignment rate between automated outputs and manual labels.

After the session, we fed the consolidated features into gpt-4o \cite{openai2024gpt4ocard} using a structured, theory-informed prompt. The LLM processed each time bucket (60s) request within 10s, generating narrative analyses that described group interactions and suggested how individual behavioral patterns might relate to specific construct facets. While the model does not produce deterministic labels, its generated interpretations demonstrated plausible and theoretically coherent mappings, often reflecting the observer’s impression of the session. This suggests the potential of LLMs to support construct-aligned reasoning from low-level multimodal signals, especially in exploratory or reflective analytics contexts.

\section{Limitation \& Future Work}
\textit{Wearable sensing}: Arduino badges may disconnect due to overheating, and calibration drift between devices affects positional accuracy. Improvements in robustness and wearability are needed.

\textit{Indicators}: Although grounded in learning theory, the current set of indicators may not fully capture the richness of constructs and may lose critical information from features. Further testing with diverse constructs is needed to assess the generalizability and adequacy of these indicators.

\textit{Real-time feedback}: Real-time LLM feedback is not yet supported; features are stored in InfluxDB and analyzed post-session. We aim to support continuous updates and live insights.

\textit{Evaluation}: Further empirical studies are needed to validate construct prediction accuracy and assess the educational impact of system feedback.

\section{Conclusion}
In summary, BadgeX demonstrates a novel synergy between wearable IoT sensing and LLM-based interpretation to support collaborative learning. By addressing technical challenges and focusing on privacy and usability, this approach can move us closer to AI-augmented classrooms where group learning processes are visible and supported in real time.

\bibliographystyle{ACM-Reference-Format}
\bibliography{sample-base}

\end{document}